# Optimal Geometric Partitions, Covers and K-Centers

MUGUREL IONUŢ ANDREICA, ELIANA-DINA TÎRŞA
Politehnica University of Bucharest
Splaiul Independentei 313, sector 6, Bucharest
ROMANIA
{mugurel.andreica,eliana.tirsa}@cs.pub.ro https://mail.cs.pub.ro/~mugurel.andreica/

CRISTINA TEODORA ANDREICA, ROMULUS ANDREICA
Commercial Academy Satu Mare
Mihai Eminescu 5, Satu Mare
ROMANIA
academiacomerciala@yahoo.com http://www.academiacomerciala.ro

MIHAI ARISTOTEL UNGUREANU
Romanian-American University
Bd. Expozitiei 1B, sector 1, Bucharest
ROMANIA
m_a_ungureanu@yahoo.com http://www.rau.ro

*Abstract:* - In this paper we present some new, practical, geometric optimization techniques for computing polygon partitions, 1D and 2D point, interval, square and rectangle covers, as well as 1D and 2D interval and rectangle K-centers. All the techniques we present have immediate applications to several cost optimization and facility location problems which are quite common in practice. The main technique employed is dynamic programming, but we also make use of efficient data structures and fast greedy algorithms.

## 1 Introduction

In this paper we present some geometric optimization techniques for computing polygon partitions, interval and rectangle K-centers and 1D and 2D point, interval, square and rectangle covers. The techniques have immediate applications to several cost optimization and facility location problems which appear quite often in practical settings. The main technique we use is dynamic programming, together with efficient data structures. For some problems, we also make use of binary search and greedy algorithms. For most of the problems we will only show how to compute the best value of the optimization function; the actual solution will be easy to compute using some auxiliary information.

The rest of this paper is structured as follows. In Section 2 we present a generic algorithm for several types of optimal polygon partitions. In Section 3 we present a generic framework for computing optimal point and interval covers in one dimension. In Section 4 we consider square and rectangle covering problems of points in two dimensions. In Section 5 we discuss the weighted interval K-center of points on a line and in Section 6 we consider the 2D rectangle K-center problem. Finally, in Section 7 we present related work and in Section 8 we conclude and discuss future work.

## 2 Optimal Polygon Partitions

In this sub-section we consider the problem of optimally partitioning a convex polygon with n vertices into K+1 parts, by drawing K ($0 \le K \le n-3$) non-crossing diagonals. Each diagonal between two vertices i and j has a weight w(i,j) (e.g. its length). The optimality of the partitioning consists in minimizing (maximizing) the total weight of the K diagonals or the largest (smallest) weight of a diagonal (the min-sum, max-sum, min-max and max-min objectives). This problem is motivated by the need to divide convex areas into several distinct parts by building walls between points on the contour of the area. A naive approach would be to use a greedy algorithm which repeatedly chooses the smallest (or largest, respectively) weight diagonal which does not cross any previously chosen diagonal. It is easy to find counter-examples for which this algorithm does not work. Another approach would be to use an exhaustive search and generate all the polygon partitions into K+1 parts. However, there is a large number of such partitions. We provide here a method for computing their number. We consider that the vertices of the polygon are numbered from 1 to n, in clockwise or anti-clockwise order. We will compute three tables: NP(i,j,0)=the number of partitions of a convex polygon with i vertices into j parts, such that no diagonal with an endpoint at the vertex labeled with 1 is drawn; NP(i,j,1)=the same as NP(i,j,0), except that at least one diagonal with an endpoint at the vertex 1 is drawn; NP(i,j,2)=NP(i,j,0)+NP(i,j,1). We have NP(i,1,1)=NP(i≤2,*,*)=0 and NP(i≥3,1,0)=1. For i≥3, j>1 we have NP(i,j,0)= NP(i-1,j,2)+NP(i-1,j-1,2). This is because if no diagonal has an endpoint at vertex 1 then the polygon is equivalent to another one having only i-1 vertices, in which vertex 1 does not exist. A second choice considers drawing the diagonal (2,i) and focusing on the remaining polygon with i-1 vertices. For NP(i,j,1) we use the following equation:

$$NP(i,j,1)=\sum_{p=3}^{i-1}\sum_{q=1}^{j-1} NP(p,q,0)\cdot NP(i-p+2,j-q,2) \quad (1)$$

In (1), p is the smallest index such that the diagonal (i,p) is drawn. NP(n,K+1,2) contains the number of partitions. For n=25 and K+1=10 there are 1,918,404,688,200 partitions, making the exhaustive search unfeasible.

We will now present a generic dynamic programming algorithm which can solve all of the problems we mentioned. We compute a table OPT(i,j,d,l)=the optimal value of the objective function if we consider only the vertices i, i+1, …, j, we draw d diagonals with both endpoints in the interval of vertices [i,j] and:

- if l=0, then the diagonal (i,j) is not one of the d diagonals drawn.
- if l=1, then the diagonal (i,j) may or may not be one of the d diagonals drawn.

We have OPT(i,j,0,0)=OPT(i,j,0,1)=0 for all the pairs (i,j). All the other entries of the table will be initialized to +∞ for a minimizing objective function and -∞ for a maximizing function. For j-i≥2 and d≥1, we have:

$$OPT(i,j,d,0)=\underset{\substack{i\le p<j\\ 0\le q\le d}}{optf}\{addf(OPT(i,p,q,1),OPT(p,j,d-q,1)\} \quad (2)$$

$$OPT(i,j,d,1)=optf\begin{cases} OPT(i,j,d,0) \\ addf(w(i,j),OPT(i,j,d-1,0)) \end{cases} \quad (3)$$

optf(x,y) is one of the min(x,y) or max(x,y) functions and addf(x,y) is either (x+y), min(x,y) or max(x,y). The pseudocode of the algorithm is presented below:

```
OptimalPolygonPartition(n, K):
initialize the OPT(i,j,k,l) table
for li=3 to n do // li=3, 4, …, n
 for i=1 to n-l+1 do // i = 1, 2, …, n-l+1
 j=i+li-1
 for d=1 to K do
   for p=i to j-1 do
    for q=0 to d do
     OPT(i,j,d,0)=optf(addf(OPT(i,p,q,1), OPT(p,j,d-q,1)),
                       OPT(i,j,d,0))
     OPT(i,j,d,1)=optf(OPT(i,j,d,0), addf(w(i,j), OPT(i,j,d-1,0))
  return OPT(1,n,K,0)
```

The result is found at OPT(1,n,K,0). Note that (1,n) is not a diagonal (although it is handled that way by the algorithm). This is why we need to choose the entry which does not contain the segment (1,n). The time complexity of the algorithm is $O(n^3 \cdot K^2)$. The min-sum objective is achieved by setting optf(x,y)=min(x,y) and addf(x,y)=(x+y). For the max-sum objective, we have optf(x,y)=max(x,y) and addf(x,y)= (x+y). For the min-max objective we set optf(x,y)=min(x,y) and addf(x,y)= max(x,y); for the max-min objective we set optf(x,y)=max(x,y) and addf(x,y)=min(x,y). We can also handle the min-min and max-max objectives, but these can be trivially solved by an algorithm that chooses the minimum weight diagonal (or the maximum weight diagonal, respectively), together with any other non-crossing K-1 diagonals. There is just one more thing to consider. When addf(x,y)=min(x,y), we need to ignore arguments which are equal to 0; if one of the arguments is equal to 0, we should just return the other argument. The algorithm we presented can be extended to any simple polygon, by computing which pairs of vertices (i,j) form a valid diagonal (one which is completely located inside the polygon and which does not cross any of the polygon's sides). Then, if (i,j) does not constitute a valid diagonal, we will set OPT(i,j,d,1) to OPT(i,j,d,0).

# 3 Covering Points and Intervals (1D)

In this section we consider collections of points and intervals located on the real line. Each point i is located at coordinate $x_i$ and has a weight $w_{pt,i}$. Each interval j is a closed interval $[l_j, h_j]$ and has a weight $w_{int,j}$. We consider that there are n points and m intervals. With this model, we can define several covering problems, solved by the following generic algorithm.

**Step 1.** Build an array p with the n coordinates of the points and the $2\cdot m$ coordinates of the intervals' endpoints. Then sort this array, such that $p_1 \le \ldots \le p_{n+2\cdot m}$. For each $p_i$, store its type $type_i$ (*point, left endpoint, right endpoint*) and, in case it is a right endpoint, store the position *lpos(i)* of its corresponding left endpoint in the array p ($p_i=h_j$ and $p_{lpos(i)}=l_j$ for some interval j); if $p_i$ is either a left or right endpoint, we store the position *lpoint(i)* in the array p of the point which is closest to $p_i$ and to its left (or 0 if no such point exists); if $p_i$ is a point, then we compute the position *lrint(i)* in the array p of the rightmost right endpoint of an interval which is located to the left of $p_i$ (or 0 if none can be found). The sorting stage will take $O((n+2\cdot m)\cdot \log(n+2\cdot m))$ time and computing the values *lpos*, *lpoint* and *lrint* takes $O(n+2\cdot m)$ time. We will also compute an array *index*, where *index(i)=j* if ($type_i$=point and $p_i=x_j$) or ($type_i$=left endpoint and $p_i=l_j$) or ($type_i$=right endpoint and $p_i=r_j$).

**Step 2.** Compute a table T, where $T_i$ is the optimal value of the covering function, considering only the positions $p_1,\ldots,p_i$. We use dynamic programming for this:

**<u>Generic Dynamic Programming Algorithm(n,m):</u>**

**for** $i=1$ **to** $n+2\cdot m$ **do**
  **if** *($type_i$=point)* **then call** $F_{point}(i)$
  **else if** (*$type_i$=right endpoint*) **then call** $F_{right_endpoint}(i)$

All that needs to be defined are the functions $F_{point}$ and $F_{right_endpoint}$. The optimal value will be found either at position i corresponding to the x coordinate of the rightmost point or at position j, corresponding to the rightmost right endpoint of an interval. We will now present several problems which use this framework.

## 3.1 Covering all the points with a subset of intervals having minimum total weight

We need to select a subset S of the m intervals such that for every point i, there exists some interval j in S, such that $l_j \le x_i \le h_j$. The weight of the subset is

$$w(S) = \sum_{j\in S} w_{int,j} \qquad (4)$$

We need an algorithm for determining a subset S with minimum total weight. We will define the two functions as follows: $F_{point}(i)=(T_i=+\infty)$ and:

**<u>$F_{right\ endpoint}$(i):</u>**

$T_i=+\infty$
$p_{first}=max\{lpoint(lpos(i)),1\}$ // the rightmost point to the left of $p_{lpos(i)}$
**for** $j=p_{first}$ **to** $i-1$ **do** // consider $p_{first}$ + all the points in $[l_{index(i)},r_{index(i)}]$
  **if** (($type_j$=point) **and** ($T_j+w_{int,index(i)}<T_i$)) **then**
    $T_i=T_j+w_{int,index(i)}$
**if** (($lpoint(i)>0$) **and** ($T_i<T_{lpoint(i)}$)) **then**
  $T_{lpoint(i)}=T_i$ // update the rightmost point to the left of $p_i$

The minimum weight of the subset is found at $T_j$, where $p_j$ corresponds to the rightmost point. The algorithm has time complexity $O((n+2\cdot m)^2)$, but we can improve it, by maintaining a stack of *(position, value)* pairs. The pairs will be sorted in increasing order of their position and of their value. In the last line of the function $F_{right_endpoint}(i)$, after updating the value of $T_{lpoint(i)}$, we will add to the stack the pair *(position=lpoint(i), value= $T_{lpoint(i)}$)*, but not before removing from the top of the stack all the pairs with a value larger than $T_{lpoint(i)}$. Then, on this stack, in the $F_{right_endpoint}(i)$ function, we will binary search the pair *(pos,val)* with the smallest position *pos* which is larger than or equal to *lpoint(lpos(i))* and set $T_i$ to $val+w_{int,index(i)}$ (or $+\infty$ if no position is found). The time complexity of the algorithm is $O((n+2\cdot m)\cdot \log(n+2\cdot m)+m\cdot\log(n))$. Instead of the stack we could use a balanced tree (AVL tree, red-black tree) or a segment tree for point updates and range minimum queries [12].

## 3.2 Covering all the intervals with a subset of points having minimum total weight

We need to select a subset S of the n points such that for every interval j, there exists some point i in S, such that $l_j \le x_i \le h_j$. The weight of the subset is

$$w(S) = \sum_{i\in S} w_{pt,i} \qquad (5)$$

We need an algorithm for determining a subset S with minimum total weight. First, we will eliminate all the intervals which fully contain inside them some other interval. In order to do this, we sort the $2\cdot m$ endpoints of the intervals (together with their type - left or right endpoint) and traverse them. We will maintain a balanced tree with the left endpoints of the currently open, unmarked, intervals. When we encounter the left endpoint $l_i$ of an interval i, we insert it into the tree. When we encounter the right endpoint $h_i$ of an unmarked interval i, we mark all the intervals j whose left endpoint $l_j$ is inside the tree and $l_j<l_i$ and afterwards remove all the $l_j$'s from the tree, as well as $l_i$. We will remove from the set every marked interval. Thus, in O(m·log(m)) time, we can determine all the intervals which contain inside themselves some other interval (if instead we mark at each step the interval j with the minimum $l_j$ in the tree, then j is certainly not contained inside any interval). From now on, we will assume that none of the m intervals contain each other. We will define the two functions, $F_{point}(i)=(T_i=w_{pt,index(i)} + (if\ (lrint(i)>0)\ then\ T_{lrint(i)}\ else\ 0))$ and $F_{right_endpoint}$:

**$F_{right_endpoint}$(i):**
*$T_i=+\infty$*
**for** *j=lpos(i)+1* **to** *i-1* **do //** consider all the points in $[l_{index(i)},r_{index(i)}]$
  **if** ((*$type_j$=point*) **and** (*$T_j<T_i$*)) **then**
    *$T_i=T_j$*

The answer is found and $T_j$, where $p_j$=the rightmost endpoint of an interval. The time complexity of the algorithm can be improved by using a double-ended queue (deque) which stores *(position, value)* pairs. Just like in the previous case, these pairs will be sorted increasingly both according to the position and to the value. At the end of the $F_{point}$(i) function, we will add to the end of the deque the pair *(position=i, value=$T_i$)*, after removing from the end of the deque all the pairs with a value which is larger than or equal to $T_i$. At the beginning of the $F_{right_endpoint}$(i) function, we will remove from the front of the deque all the pairs *(pos,val)* for which *pos<lpos(i)+1*. After that, we will select the pair *($pos_f$, $val_f$)* remaining at the front of the deque (if the deque is not empty) and set $T_i$ to $val_f$; if the deque is empty, then $T_i=+\infty$. The time complexity of this stage is O(n+2·m), but the overall complexity of the algorithm is dominated by the sorting stage (if the intervals and the points are already sorted, then the overall complexity becomes linear). Instead of the deque, we can use balanced trees or a segment tree built on the points $\{p_i\}$, with point updates and range minimum queries [12], but the time complexity of the dynamic programming step would become O((n+2·m)·log(n+2·m)).

### 3.3 Min-Max and Max-Min Weighted Covers

If we are interested in covering all the points with a subset S of intervals in which the maximum weight of an interval in S is minimized, we will perform a binary search in order to find the minimum maximum weight. Let's assume that we chose the value W in the binary search procedure. We now need to perform a feasibility test. We will remove all the intervals having a weight larger than W and then verify whether the remaining intervals cover all the n points. We can perform the test in O(n+m) time, by traversing from left to right the endpoints of the intervals and the points and maintaining a counter *nopen* with the number of "open" intervals (we increment it by 1 at every left endpoint and decrement it by 1 at every right endpoint we encounter). Then, for each point, we just verify that *nopen* is larger than 0. Note that the points and the intervals' endpoints only need to be sorted once (in the beginning). We can use the same algorithm for a max-min objective function (just remove the intervals with a weight smaller than W). The same technique can be used for min-max and max-min coverings of the intervals with points (this time we remove the points with weights larger or smaller than the value tested in the binary search). The feasibility test consists of maintaining the rightmost not-removed point encountered during the left-to-right traversal; then, during the traversal, for each right endpoint of an interval I, we verify whether the rightmost (so far) not-removed point is located to the right of the left endpoint of I.

## 4 Covering Points in the Plane (2D)

In this section we consider two point covering problems in two dimensions. We are given n points in the plane; point i is located at coordinates $(x_i, y_i)$. We will present efficient algorithms for covering all the points by K≤3 axis-aligned rectangles, with the min-sum or min-max area objective. We also consider the case of fixed size rectangles covering a maximum weight subset of points.

### 4.1 Covering all the points with K≤3 rectangles

We want to cover all the n points using K≤3 axis-aligned rectangles, subject to an area minimization criterion. We will compute the minimum and maximum x- and y-coordinates of the points: $x_{min}$, $y_{min}$, $x_{max}$ and $y_{max}$. If K=1, the rectangle's area is $(x_{max}-x_{min})\cdot(y_{max}-y_{min})$. If we are using squares instead of rectangles, the area will be $(\max\{x_{max}-x_{min}, y_{max}-y_{min}\})^2$. For K≥2, because there are at most 3 rectangles and the minimum bounding rectangle (MBR) of the n points has 4 sides, the first rectangle will have two of its sides along at least two sides of the MBR. We will consider first all the choices of two adjacent sides (left and up, left and down, right and up, right and down). Each choice determines a corner of the rectangle (upper left, lower left, upper right or lower right). We will now attempt all the $O(n^2)$ possibilities for the horizontal length $L_x$ and vertical length $L_y$ of the rectangle (these lengths are determined by the fixed corner and the x- and y-coordinates of some points). Once the rectangle is fixed, we remove all the points contained inside it and repeat the same procedure for K-1 rectangles. For $K \geq 2$, we also have to consider the case when the two sides of the first rectangle are parallel to each other (up and down or left and right). We will discuss the up and down case only, as the other one is similar (by exchanging the x and y coordinates of the points, the left-right case becomes the up-down case). We consider the points sorted in ascending order of their x-coordinates: $x_1 \leq x_2 \leq \ldots \leq x_n$. The first rectangle will be the center rectangle and will have a rectangle to its left and another one to its right. In O(n) time we will compute the following values: $yc_{l,max}(i) = \max\{y_1,\ldots,y_i\}$, $yc_{l,min}(i)=\min\{y_1,\ldots,y_i\}$. In a similar manner, we compute $yc_{r,max}(i)=\max\{y_i, y_{i+1}, \ldots,y_n\}$ and $yc_{r,min}(i)= \min\{y_i, y_{i+1}, \ldots,y_n\}$. In the up-down case, the vertical length of the center rectangle is equal to $L_y=y_{max}-y_{min}$. We will now attempt all the $O(n^2)$ possibilities for the x-coordinates of the left and right sides of the center rectangle. When the left side is at coordinate $x_i$, and the right side at coordinate $x_j$, the area of the center rectangle will be $A_{center}=L_y\cdot(x_j-x_i)$, that of the left rectangle will be $A_{left,i}=(x_{i-1}-x_1)\cdot(yc_{l,max}(i-1)-yc_{l,min}(i-1))$ and that of the right rectangle will be $A_{right,j} = (x_n-x_{j+1})\cdot(yc_{r,max}(j+1)-yc_{r,min}(j+1))$. We will choose the pair (i,j) for which $\max\{A_{center}, A_{left,i}, A_{right,j}\}$ is minimum (for the min-max area objective) or $(A_{center}+A_{left,i}+A_{right,j})$ is minimum (for the min-sum area objective). The time complexity of the algorithm is T(n,K), where T(n,1)=O(n) and $T(n,K\geq 2)= 4\cdot O(n^2)\cdot T(n,K-1)+2\cdot O(n^2)=O(n^{2\cdot K-1})$ (the constants 4 and 2 stand for the total of 6 pairs of sides of the MBR). If we are using squares instead of rectangles, the time complexity for the min-sum objective can be reduced to $T(n,K)=O(n^K)$ (because we only need to choose the length of one side of the square or the x-coordinate of its left side, instead of two sides or two x-coordinates for a rectangle) and the min-max area objective can be solved in O(n) time [7]. We can improve the time complexity for K=2, by storing the points into several orthogonal range search data structures (like range trees), where the weight of each points is its x- (y-) coordinate. Then, after choosing the position of the first rectangle, we can find in $O(\log^2(n))$ time the position of the last rectangle (by querying the range trees with the remaining area, which is the union of two rectangles, and obtaining the extreme coordinates of the remaining points). Thus, For $2 \leq K \leq 3$, we have $T(n,K)=O(n^{2\cdot K-2}\cdot\log^2(n))$. For squares, the complexity improves to $O(n^{K-1}\cdot\log^2(n))$. When the points belong to an mxm grid and $n=O(m^2)$, we have only $O(m^2)$ choices for the side lengths of a rectangle (O(m) for squares) having two sides aligned with two sides of the MBR. Thus, $T(m,1)=O(m^2)$ and $T(m,K)=O(m^2)\cdot T(m,K-1)=O(m^{2\cdot K})$. For $K \geq 2$, we can optimize T(m,K) to $O(m^{2\cdot K-2})$, by having $T(m,2)=O(m^2)$. Let's assume that the first rectangle has its upper right corner fixed at $(x_{max}, y_{max})$ and that we selected the side lengths $L_x$ and $L_y$ for it. We will precompute in $O(m^2)$ time several tables: $x_{t,min}(i,j)$=the minimum value of the x-coordinate of a point inside the rectangle (0,0)-(i,j), $x_{t,max}(i,j)$, $y_{t,min}(i,j)$ and $y_{t,max}(i,j)$ (having obvious and analogous meanings). If there is a point at coordinates (i,j) then $x_{t,min}(i)=\min\{x_{t,min}(i-1,j), x_{t,min}(i,j-1),i\}$; otherwise, i is not a candidate for $x_{t,min}(i,j)$. After selecting the lengths $L_x$ and $L_y$, all the points in the rectangle $(x_{max}-L_x, y_{max}-L_y)-(x_{max},y_{max})$ are removed. We now need to find the extreme x- and y-coordinates of all the remaining points. We perform this step in O(1) time. The minimum x-coordinate of one of the remaining points is $\min\{x_{t,min}(x_{max},y_{max}-L_y-1),x_{t,min}(x_{max}-L_x-1,y_{max})\}$. The other extreme x- and y-coordinates can be found in O(1), too. The cases when other corners are fixed are similar. For squares we get $T(m,K\leq 2)=O(m^2)$ and $T(m,3)=O(m^3)$.

## 4.2 Covering a maximum weighted subset of points with K≤3 rectangles of fixed size

We are given n points in the plane. Each point i is located at coordinates $(x_i,y_i)$ and has a weight $w_i$. We want to place $K \leq 3$ identical rectangles of sizes $L_x$ (horizontal size) and $L_y$ (vertical size) at disjoint positions in such a way that a subset of points with maximum total weight is covered by them. We will start with the K=1 case. We sort the points in increasing order of their x-coordinates and build a segment tree on their sorted y-coordinates (the y-order): $y_1 \leq y_2 \leq \ldots \leq y_n$. The segment tree will be used for range addition updates and range maximum queries [12]. For each i, we compute plow(i)=the smallest index such that $y_i-y_{plow(i)} \leq L_y$. We will sweep the points from left to right with a vertical region of infinite height and horizontal length $L_x$. We have two kinds of events: a point $(x_p, y_p)$ enters the vertical region or leaves the region (p is its index in the y-order). When the point enters the region, we range update the interval [plow(p),p] in the segment tree with the value $w_p$; when it leaves, we update the same interval with the value $-w_p$. After each update, we query the interval [1,n] and obtain the maximum weight of a subset of points covered by a rectangle whose left and right sides are those of the vertical region and whose lower side is at one of the coordinates $y_1,y_2,\ldots,y_n$. Because there are O(n) events (which can be computed in O(1) time per event), the algorithm's complexity is $O(n\cdot\log(n))$. We can extend this algorithm to K=2 rectangles and maintain the same complexity. In this case, the rectangles will be separated by a line (horizontal or vertical). We will present the vertical case, as the horizontal one is handled in a similar manner. From left to right, we compute $maxw_l[i]$=the maximum weight of a subset of points covered by a rectangle, considering only the points 1,2,…,i (sorted according to their x-coordinates). Similarly, we compute $maxw_r[i]$ for the points i, i+1, …, n (moving from right to left). The maximum weight of a subset covered by both rectangles is $\max\{maxw_l[i]+maxw_r[i+1]\}$, $0 \leq i \leq n$. For K=3 rectangles, we separate the third rectangle by the other two using one of the O(n) vertical and horizontal lines and solve the cases K=1 and K=2 for the corresponding half-planes. The time complexity becomes $O(n^2\cdot\log(n))$. If the points belong to an mxm grid an $n=O(m^2)$, we can optimize the complexity. We will compute a table *wsum(i,j)*=the sum of the weights of the points inside the rectangle (0,0)-(i,j). Using this table, we compute four others: *wmax(i,j,q)*=the maximum weight of a subset of points covered by a rectangle, considering only the points in the quadrant q ($1 \leq q \leq 4$), with regard to the point (i,j). We can compute all of these tables in $O(m^2)$ time. Afterwards, we test all the $O(m^2)$ pairs of separating lines ($l_1$, $l_2$) ($l_1$ separates the third rectangle from the first two and $l_2$ separates the first two rectangles) and obtain the maximum weight of the points covered in O(1) time.

# 5 Interval K-Centers of Points (1D)

We are given n points located on the real line. Each point i is located at coordinate $x_i$ and has a weight $w_i$. We are interested in placing (at most) K intervals of a fixed length L, such that the maximum weighted distance from a point to an interval is minimized. The distance from a point $x_j$ to an interval [a,b] is: 0, if $a \leq x_j \leq b$ ; $w_j \cdot (a-x_j)$, if $x_j < a$ ; $w_j \cdot (x_j-b)$, if $x_j > b$. This problem is closely related to the weighted K-center problem on trees, in which the centers are points. We will solve the problem by binary searching the maximum weighted distance D. We will then perform a feasibility test, in order to verify if D is larger than (or equal to) the optimal distance. For each point i, we will compute $d_i = D/w_i$, the maximum allowed (non-weighted) distance on the real line between $x_i$ and a point of an interval. An interval [a,b] of fixed length L is properly defined by its left endpoint a (the interval is [a, a+L]). Thus, for each point i, the left endpoint of one of the K intervals must be located somewhere in the interval $[x_i-d_i-L, x_i+d_i]$. The problem is equivalent to finding (at most) K points, such that every interval $[lx_i=x_i-d_i-L, rx_i=x_i+d_i]$ contains at least one of the points. We will now solve an optimization problem: compute the minimum number P of points required such that every interval $[lx_i, rx_i]$ contains at least one point. This is the *Hitting Set Problem on a Line*, which can be solved by a standard $O(n \cdot \log(n))$ greedy algorithm. We sort the $[lx_i, rx_i]$ intervals according to their right endpoint. Then we traverse them from left to right and place a new point at the right endpoint of every interval not containing the last point placed so far. If the number of points placed P is larger than or equal to K, then we will test a smaller value of D; otherwise, we test a larger value.

For K=1 we can solve the problem in O(n) time, if the points are sorted. For each point i we define a half-line with slope $w_i$, starting at $(x_i,0)$. If the left endpoint of the interval is placed at x=l, then we need to find the half-line j in the set $\{i | x_i \leq l\}$ with maximum y-coordinate. With a deque, we can update and retrieve the maximum in O(1) amortized time when moving from left to right (we can do something similar for the right endpoint when moving from right to left). By storing these values for each of the O(n) candidate left and right endpoints and testing all of them, we obtain an O(n) algorithm.

We will now extend the K-center problem as follows. We are also given the positions of P centers and want to place K *extra* centers, such that the maximum weighted distance from a point to the closest center (out of the P+K centers) is minimum. We call this extension the (K+P)-center problem and we will solve it by slightly modifying the solution for the K-center problem. During the binary search, after assigning an interval $[lx_i, rx_i]$ to every point i, we will verify if any of these intervals contains one of the left endpoints of the P centers which are fixed. We will ignore all such intervals in the decision algorithm, because the points associated with these intervals are "satisfied" by the P fixed centers (i.e. are within weighted distance D from one of these centers and do not require the placement of a new center just for them; D is the weighted distance considered in the binary search). In order to exclude the points which are satisfied by the P fixed centers, we sort the intervals assigned to each point and the left endpoints of the fixed centers and maintain a set S of open intervals (an interval is opened when its left endpoint is encountered and closed when we encounter its right endpoint). When encountering the left endpoint of a fixed center, we mark all the open intervals at that moment. We note that every K-center problem which can be solved by binary searching the maximum (weighted) distance from a point to its closest center can be extended similarly to the (K+P)-center problem.

# 6 Rectangle K-Centers (2D)

We are given n (unweighted) points in the plane; point i is located at coordinates $(x_i, y_i)$. We want to place $K \leq 3$ rectangles with side lengths $L_x$ (horizontal side) and $L_y$ (vertical side) such that the maximum ($L_\infty$) distance from a point to the rectangle is minimized. We binary search the maximum distance D and set $L_x' = L_x + 2 \cdot D$ and $L_y' = L_y + 2 \cdot D$. The feasibility test consists of verifying if all the points can be covered by K rectangles of sizes $L_x'$ and $L_y'$. We use the ideas from Subsection 4.1. One of the rectangles must have two sides aligned with two sides of the MBR. If we choose two opposite sides of the MBR, the test can be performed in O(n) time. If we choose two adjacent sides, the time complexity of the test is $T(n,K>1)=4 \cdot T(n,K-1)=O(n)$, with $T(n,1)=O(n)$.

The corresponding (K+P)-center problem is solved by inflating accordingly the P fixed rectangles and ignoring in the decision problem all the points which are contained in at least one of the P inflated rectangles. We can easily find all the points contained in at least one of the P inflated rectangles in $O(n \cdot P)$ time, but, if P is large, we can do this in $O(P \cdot \log(n) + n \cdot \log(n))$. We construct a (dynamic) range tree with all the n points. Then, for each of the P rectangles, we report in O(log(n)+q) time, all the q points contained in it; afterwards, we remove the q points from the range tree (in order to avoid reporting the same point multiple times, which would again take O(n·P time).

When the points have different weights, the rectangle K-center problem is equivalent to deciding if a set of n rectangles is K-pierceable.

# 7 Related Work

Minimum length partitions of polygons with n vertices into parts having at most m edges were considered in [1]. Polygon triangulations are a special case of partitions and were discussed in many papers, like [2,3]. A closed formula for the total number of polygon partitions was given in [13]. One-dimensional coverings of intervals with points are related to the Hitting Set and P-coverage problems [4,5]. Two-dimensional coverings of points with two and three squares [6,7] and rectangles [8] have been studied extensively and efficient algorithms were proposed. One-dimensional K-centers have been investigated for a long time, but they still present significant interest [9], as well as their (weighted) two-dimensional versions [10,11]. Rectilinear p-piercing problems [10] are closely related to the interval and rectangle K-center problems we discussed in this paper.

# 8 Conclusions and Future Work

In this paper we presented several geometric optimization techniques for computing optimal polygon partitions, 1D and 2D point, interval, square and rectangle covers, 1D weighted interval K-centers and 2-D rectangle K-centers (only in some special circumstances). All the techniques we presented are efficient and can be easily used in a practical setting. All the results we presented are new, but some of them are based greatly on standard algorithms and data structures or are only extensions of techniques presented in other papers. As future work, we intend to tackle several more difficult problems, like optimal area polygon partitions, covering points with more than 3 squares or rectangles (and with different optimization objectives) and weighted rectangle K-centers. We also intend to study the interval and rectangle K-median problem. We believe that some of the techniques that we developed in this paper can be (non-trivially) extended to handle more difficult cases. All the problems we studied are motivated by practical situations, like minimum cost space division and optimal facility locations (with respect to some optimization criterion), like hospitals, public institutions, processing plants, oil extraction platforms and real estates.